\documentclass[twocolumn,prb]{revtex4}
\usepackage{amsfonts}
\usepackage[T1]{fontenc}
\usepackage{amsmath,amsbsy,amssymb,graphicx}
\usepackage{times}

\begin{document}

\title{{\Large Valleytronics on the Surface of Topological Crystalline
Insulator:}\\
{\Large Elliptic Dichroism and Valley-Selective Optical Pumping} }
\author{Motohiko Ezawa}
\affiliation{Department of Applied Physics, University of Tokyo, Hongo 7-3-1, 113-8656,
Japan }

\begin{abstract}
The low-energy theory of the surface of the topological crystalline
insulator (TCI) is characterized by four Dirac cones anisotropic into the $x$
and $y$ directions. Recent experiments have shown that the band gap can be
introduced in these Dirac cones by crystal distortion by applying strain to
the crystal structure. The TCI surface provides us with a new way to
valleytronics when gaps are given to Dirac cones. Indeed the system has the
Chern number and three valley-Chern numbers. We investigate the optical
absorption on the TCI surface. It shows a strong elliptic dichroism though
the four Dirac cones have the same chiralities. Namely, it is found that the
absorptions of the right- and left-polarized light are different, depending
on the sign of mass and the location of the Dirac cones, owing to the
anisotropy of the Dirac cone. By measuring this elliptic dichroism it is
possible to determine the anisotropy of a Dirac cone experimentally.
\end{abstract}

\maketitle

%\date{}

\address{{\normalsize Department of Applied Physics, University of Tokyo, Hongo
7-3-1, 113-8656, Japan }}

\section{Introduction}

Valleytronics is a promising candidate of the next generation electronics%
\cite{Rycerz,Xiao07,Akh,Yao08,VMetal,Kirch,Xiao}. It is a technology of
manipulating the degree of freedom to which inequivalent degenerate state an
electron belongs near the Fermi level. The main target of valleytronics is
the honeycomb lattice system such as graphene. Indeed, the honeycomb
structure is an ideal play ground of valleytronics since it has two
inequivalent Dirac cones or \textit{valleys}. A key progress in
valleytronics is valley-selective optical pumping\cite%
{Yao08,Xiao,Li,EzawaOpt,EzawaSV,Stille,Carbotte}. By applying circular
polarized light in a gapped Dirac system, we can selectively excite
electrons in one valley based on the property that two valleys have opposite
chiralities. It is known as the circular dichroism. Valley-selective pumping
has been observed\cite{Mak,Mak2,Spl,Zeng,Cao,Wu} in the transition-metal
dichalcogenides such as MoS$_{2}$, where there exists a direct gap between
the conduction and valence bands for Dirac fermions.

However, the valleytronics is not restricted to the honeycomb system.
Recently, the topological crystalline insulator (TCI) attracts much
attention due to its experimental realizations\cite{Ando,Xu,Dz} in Pb$_{1-x}$%
Sn$_{x}$Te. It is a topological insulator protected by the mirror symmetry%
\cite{Fu,Hsieh}. The remarkable properties of the TCI is that there emerge
four topological protected surface Dirac cones, as has been observed in the
angle-resolved photoelectron spectroscopy (ARPES) experiment\cite{Ando,Xu,Dz}%
. The appearance of several topologically protected Dirac cones enables us
to use the TCI as the basic material for the valleytronics. Recent
experiments\cite{Okada} show that the band gap can be introduced in the
surface Dirac cones by crystal distortion by applying strain to the crystal
structure.

In this paper, we investigate the optical absorption of the TCI surface. The
key properties of surface Dirac cones are that all of them have the same
chirality but that each of them has a particular anisotropy. Based on the
anisotropy, we can selectively excite electrons in different valley by the
elliptically polarized light. This is a new type of dichroism different from
the circular dichroism. We call it an \textit{elliptic dichroism}. We
propose an experimental method to determine the anisotropy of the velocities
and the band gap of Dirac cones with the use of elliptic dichroism. Our
finding will open a new way of the valleytronics based on the TCI.

The present paper is composed as follows. In Section II, we introduce the
low-energy Hamiltonians $H_{X}$ and $H_{Y}$ valid near the $X$ and $Y$
points for the [001] surface, which are related by the $C_{4}$ discrete
rotation symmetry. The Hamiltonian contains the pseudospin degree of freedom
representing the cation and the anion. The $X$ ($Y$) point is separated into
a pair of the $\Lambda _{X}$ and $\Lambda _{X}^{\prime }$ ($\Lambda _{Y}$
and $\Lambda _{Y}^{\prime }$) points due to the spin-pseudospin mixing. We
then derive the four low-energy Hamiltonians describing four Dirac cones at
the $\Lambda _{X},\Lambda _{X}^{\prime },\Lambda _{Y}$ and $\Lambda
_{Y}^{\prime }$ points. They have in general Dirac electrons with different
masses $m_{X},m_{X}^{\prime },m_{Y}$ and $m_{Y}^{\prime }$. In Section III,
we study the spin and psuedospin structures around the $X$ and $Y$ points.
In Section IV, we analyze the Chern number for each Dirac cone. It is simply
given by $\pm \frac{1}{2}$ depending on the sign of the Dirac mass. Since
there are four Dirac cones, there arise the Chern number and three
valley-Chern numbers. The Chern number is a genuine topological number,
while valley-Chern numbers are symmetry-protected topological numbers. When
the mass is induced by the strain, the Chern number is zero because of the
time-reversal symmetry. On the other hand, when the mass is induced by the
exchange effect, the Chern number is $\pm 2$ per surface. In Section V, we
investigate optical absorption and elliptic dichroism by exciting massive
Dirac electrons by the right or left elliptically polarized light. We show
that the optical absorption is determined by the Chern number of each Dirac
cone and that the elliptic dichroism occurs owing to the anisotropy of a
Dirac cone. It is interesting that the elliptic dichroism is observable on
the surface of the TCI with the Dirac mass being induced by the strain.

\section{Hamiltonian}

\begin{figure}[t]
\centerline{\includegraphics[width=0.5\textwidth]{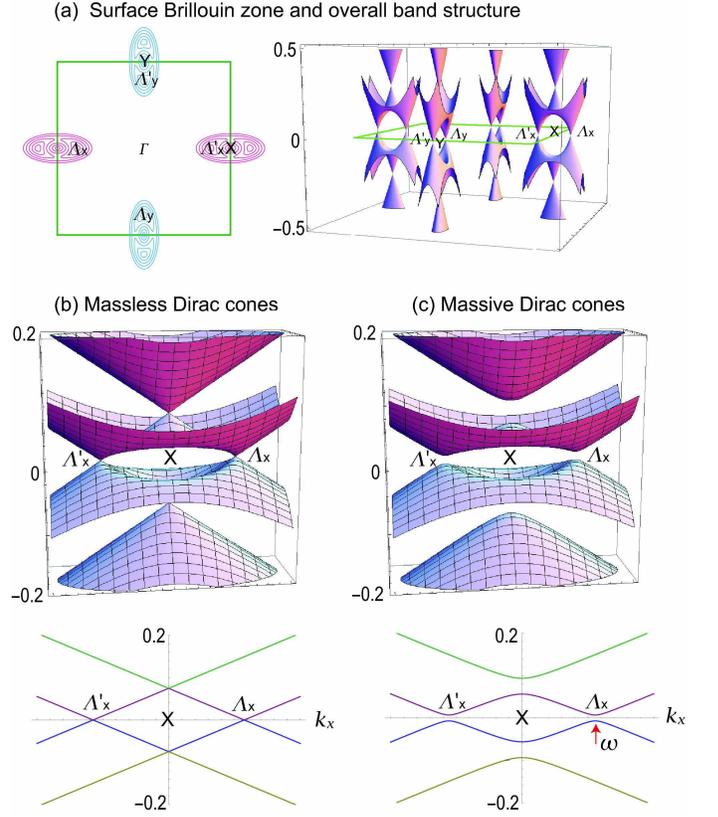}}
\caption{(Color online) (a) Surface Brillouin zone centered at the $\Gamma $
point and bounded by the $X$ and $Y$ points. There are low-energy Dirac
cones at the $\Lambda _{X},\Lambda _{X}^{\prime },\Lambda _{Y},\Lambda
_{Y}^{\prime }$ points, and high-energy Dirac cones at the $X$ and $Y$
points. (b) Detailed band structure in the vicinity of the $X$ point. Two
low-energy Dirac cones are formed at the $\Lambda _{X}$ and $\Lambda
_{X}^{\prime }$ points. (c) The gaps open when the mass term is present.}
\label{FigBand}
\end{figure}

Recent ARPES experiments\cite{Ando,Xu,Dz} show that there are four Dirac
cones at $\Lambda _{X},\Lambda _{X}^{\prime },\Lambda _{Y}$ and $\Lambda
_{Y}^{\prime }$ points in the [001] surface state of the TCI, whose band
structure we show in Fig.\ref{FigBand}(a). They may be used as the valley
degree of freedom. Two Dirac cones are present at the $\Lambda _{X}$ and $%
\Lambda _{X}^{\prime }$ points near the $X$ point but slightly away from the
$X$ point along the $x$ axis in the momentum space. The other two Dirac
cones are present at the $\Lambda _{Y}$ and $\Lambda _{Y}^{\prime }$ points
near the $Y$ point along the $y$ axis. It is notable that the Dirac cones
reside at the mirror symmetry invariant points along the $\Gamma X$ and $%
\Gamma Y$ lines rather than at the time-reversal symmetry invariant $X$ and $%
Y$ points, implying that the protected symmetry is the mirror symmetry and
not the time-reversal symmetry.

The Hamiltonian for the [001] surface states of the TCI near the $Y$ point
has been given in literature\cite{LiuFu,Fang,Wang,LiuFu04} as%
\begin{equation}
H_{Y}(\mathbf{k})=v_{2}k_{x}\sigma _{y}-v_{1}k_{y}\sigma _{x}+n\tau
_{x}+n^{\prime }\sigma _{x}\tau _{y}+m\sigma _{z}.  \label{HamilY}
\end{equation}%
The Hamiltonian near the $X$ point is given by%
\begin{equation}
H_{X}(\mathbf{k})=v_{1}k_{x}\sigma _{y}-v_{2}k_{y}\sigma _{x}+n\tau
_{x}+n^{\prime }\sigma _{y}\tau _{y}+m\sigma _{z},  \label{HamilTCI}
\end{equation}%
as we shall soon see. Here, $\mathbf{\sigma }$ and $\mathbf{\tau }$ are the
Pauli matrixes for the spin and the pseudospin representing\ the
cation-anion degree of freedom, respectively: $n$ and $n^{\prime }$ describe
the pseudospin mixing. We have set $\hbar =1$ for simplicity. Typical values
are $v_{1}=1.3$eV, $v_{2}=2.4$eV, $n=70$meV and $n^{\prime }=26$meV\cite%
{Hsieh,LiuFu04}. The term $m\sigma _{z}$\ represents the exchange
magnetization with the exchange field $m$, and acts as the mass term. It may
regarded as the Zeeman term without external magnetic field. It may arise
due to proximity coupling to a ferromagnet, as enhances the exchange
interaction to align the spin direction. We show the band structure without
and with this term in Fig.\ref{FigBand}(b) and (c), respectively.

The crystal structure of the Pb$_{1-x}$Sn$_{x}$Te is a rocksalt structure.
Accordingly, the [001] surface has the inverse $C_{4}$ discrete rotation
such that
\begin{equation}
\sigma _{x}\mapsto \sigma _{y},\qquad \sigma _{y}\mapsto -\sigma _{x}
\end{equation}%
together with
\begin{equation}
k_{x}\mapsto k_{y},\qquad k_{y}\mapsto -k_{x}.  \label{K2K}
\end{equation}%
Using this transformation, we obtain Eq.(\ref{HamilTCI}) valid near the $X$
point from Eq.(\ref{HamilY}) valid near the $Y$ point. Note that the
velocities into the $x$ and $y$ directions are different at the $Y$ point
from those at the $X$ point, as is a manifestation of the four-fold rotation
symmetry.

It follows from (\ref{HamilTCI}) that the energy spectrum is given by%
%\begin{widetext}
\begin{equation}
E(\mathbf{k})=\pm \sqrt{f\pm 2\sqrt{g}}  \label{EqEne}
\end{equation}%
%\end{widetext}
in the vicinity of the $X$ point with
\begin{subequations}
\begin{align}
f& =n^{2}+n^{\prime 2}+v_{1}^{2}k_{x}^{2}+v_{2}^{2}k_{y}^{2}+m^{2}, \\
g& =(n^{2}+n^{\prime
2})v_{1}^{2}k_{x}^{2}+n^{2}v_{2}^{2}k_{y}^{2}+n^{2}m^{2}.
\end{align}%
The band structure is shown in Fig.\ref{FigBand}. The gap closes at the two
points $(k_{x},k_{y})=(\pm \Lambda ,0)$ with $\Lambda =\sqrt{n^{2}+n^{\prime
2}}/v_{1}$ without the mass term ($m=0$). They are the $\Lambda _{X}$ and $%
\Lambda _{X}^{\prime }$ points.
\end{subequations}

\begin{figure}[t]
\centerline{\includegraphics[width=0.5\textwidth]{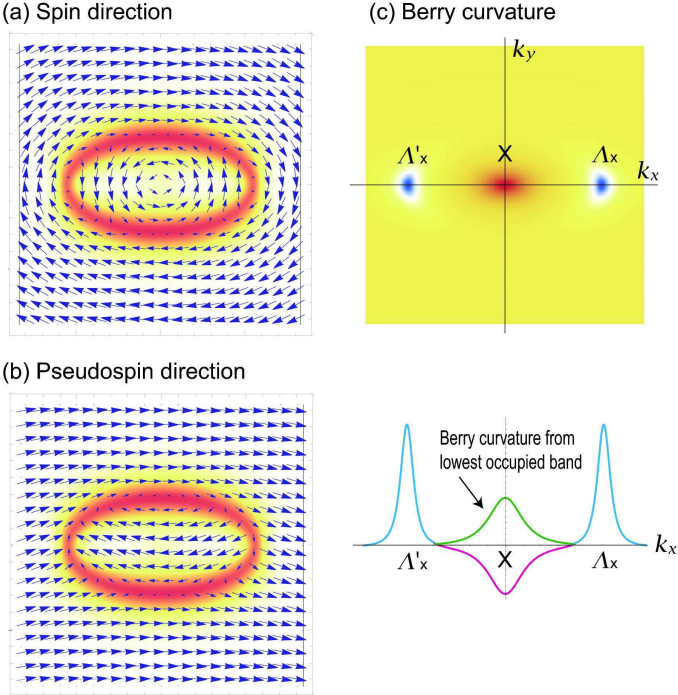}}
\caption{(Color online) (a) Spin direction of the TCI surface in the
vicinity of the $X$ point. The red oval indicates the region where the
magnitude of spin is quite small. The spin directions are opposite inside
and outside the oval. The spin rotation is clockwise (anti-clockwise) in the
low-energy (high-energy) Dirac cones at the $\Lambda _{X}$ and $\Lambda
_{X}^{\prime }$ points (the $X$ point). (b) Pseudospin direction of the TCI
surface in the vicinity of the $X$ point. The red oval indicates the region
where the magnitude of pseudospin is quite small. The pseudospin directions
are opposite inside and outside the oval. (c) Berry curvature of the highest
occupied band. It has a sharp peak (red) at the $X$ point and sharp peaks
(blue) at the $\Lambda _{X}$ and $\Lambda _{X}^{\prime }$ points. The Chern
number contribution from the Berry curvature at the $X$ point is exactly
cancelled out by the one (green) from the Dirac cone in the lowest occupied
band at the $X$ point.}
\label{SpinDirection}
\end{figure}

An intriguing feature of the TCI surface is the mass acquisition\cite%
{Hsieh,FangBer} by crystal distortion, as has been observed in recent
experiments\cite{Okada}. They are $\pm \Delta m_{X}$ and $\pm \Delta m_{Y}$
at the $\Lambda _{X}$ ($\Lambda _{X}^{\prime }$) and $\Lambda _{Y}$ ($%
\Lambda _{Y}^{\prime }$) points, respectively. Combining the mass $m$ due to
the exchange effect, the mass reads\cite{FangBer}
\begin{align}
m_{X} =&m+\Delta m_{X},\quad m_{X}^{\prime }=m-\Delta m_{X},  \notag \\
m_{Y} =&m+\Delta m_{Y},\quad m_{Y}^{\prime }=m-\Delta m_{Y}
\end{align}%
at each Dirac point. There might be other mechanisms to generate the mass.
The mass term is necessary for the valley-selective optical absorption to
occur. However the following analysis is independent of detailed origins of
the mass term.

By linearizing the band structure around the $\Lambda _{X}$ point, we obtain
the two-component low-energy Hamiltonian for massive Dirac fermions\cite%
{LiuFu,LiuFu04},
\begin{equation}
H_{\Lambda _{X}}(\widetilde{\mathbf{k}})=\tilde{v}_{1}\widetilde{k}%
_{x}\sigma _{y}-\tilde{v}_{2}\widetilde{k}_{y}\sigma _{x}+\tilde{m}%
_{X}\sigma _{z},  \label{low}
\end{equation}%
which describes physics near the Fermi level, where $\widetilde{k}_{x}=k_{x}%
\mathbf{-}\Lambda $ and $\widetilde{k}_{y}=k_{y}$, with the renormalized
velocity,
\begin{subequations}
\begin{align}
\tilde{v}_{1}=& v_{1}\sqrt{1-\frac{m_{X}^{2}n^{2}\left( n^{2}+n^{\prime
2}\right) }{\left[ \left( n^{2}+n^{\prime 2}\right) ^{2}+m_{X}^{2}n^{2}%
\right] ^{3/2}}}\simeq v_{1}, \\
\tilde{v}_{2}=& v_{2}\sqrt{1-\frac{n^{2}}{\sqrt{\left( n^{2}+n^{\prime
2}\right) ^{2}+m_{X}^{2}n^{2}}}}  \notag \\
& \simeq v_{2}n^{\prime }/\sqrt{n^{2}+n^{\prime 2}}=0.84\text{eV,}
\end{align}%
\end{subequations}
and the renormalized mass,
\begin{equation}
\tilde{m}_{X}=\text{sgn}\left( m_{X}\right) \sqrt{m_{X}^{2}+2n^{2}+2n^{%
\prime 2}-2\sqrt{(n^{2}+n^{\prime 2})^{2}+m_{X}^{2}n^{2}}}.
\end{equation}%
The energy spectrum reads%
\begin{equation}
E_{\Lambda _{X}}=\pm \sqrt{\tilde{v}_{1}^{2}\widetilde{k}_{x}^{2}+\tilde{v}%
_{2}^{2}\widetilde{k}_{y}^{2}+\tilde{m}_{X}^{2}}.  \label{ELow}
\end{equation}%
The linearized Hamiltonian around the $\Lambda _{X}^{\prime }$ point has
precisely the same expression as (\ref{low}) except that $\tilde{m}_{X}$ is
replaced by $\tilde{m}_{X}^{\prime }$. In the same way we have the
low-energy Hamiltonian around the $\Lambda _{Y}$ point,%
\begin{equation}
H_{\Lambda _{Y}}(\widetilde{\mathbf{k}})=\tilde{v}_{2}\widetilde{k}%
_{x}\sigma _{y}-\tilde{v}_{1}\widetilde{k}_{y}\sigma _{x}+\tilde{m}%
_{Y}\sigma _{z},  \label{lowY}
\end{equation}%
where $\widetilde{k}_{x}=k_{x}$ and $\widetilde{k}_{y}=k_{y}\mathbf{-}%
\Lambda $, and the similar one around the $\Lambda _{Y}^{\prime }$ point.

\section{Spin direction}

We illustrate the expectation value of the spin $\langle \mathbf{s}\rangle
=\langle \psi |\mathbf{s}|\psi \rangle $ in the vicinity of the $X$ point in
Fig.\ref{SpinDirection}(a). There is one up-pointing vortex with
anti-clockwise vorticity at the $X$ point, and there are two down-pointing
vortices with clockwise vorticity at the $\Lambda _{X}$ and $\Lambda
_{X}^{\prime }$ points\cite{Wang,Wojek,Safaei}. They describe the spin
directions of electrons in one Dirac cone at the $X$ point, and two Dirac
cones at the $\Lambda _{X}$ and $\Lambda _{X}^{\prime }$ points in Fig.\ref%
{FigBand}.

This structure is understood as follows. Let us assume $n=0$ and $n^{\prime
}=0$ in Eq.(\ref{HamilTCI}). Then the two Dirac cones in the conduction and
valence bands touch each other at the Fermi level. The effect of the term $%
n\tau _{x}$ is to shift these Dirac cones to intersect one another, forming
an intersection oval. (It is an oval and not a circle since $v_{1}\neq v_{2}$%
.) These two Dirac cones have opposite chiralities, which leads to the
opposite spin rotations inside and outside the oval. We now switch on $%
n^{\prime }$. Then the level crossing turns into the level anticrossing with
the resulting band structure as in Fig.\ref{FigBand}(a), where Dirac cones
emerge at the $\Lambda _{X}$ and $\Lambda _{X}^{\prime }$ points. The spin
rotates around each Dirac cone. The magnitude of spin, $\mathbf{s}%
^{2}=s_{x}^{2}+s_{y}^{2}+s_{z}^{2}$, is found to be quite small around the
oval [Fig.\ref{SpinDirection}(a)]. We clearly see the directions of the spin
rotation are identical in the four valleys at $\Lambda _{X}$, $\Lambda
_{X}^{\prime }$, $\Lambda _{Y}$ and $\Lambda _{Y}^{\prime }$, which
manifests the identical chirality of the four low-energy Dirac cones. On the
other hand, the spin rotation in the two high-energy Dirac cones at the $X$
and $Y$ points is opposite to the one in the low-energy Dirac cones. The
spin direction has been observed by means of spin-resolved ARPES\cite%
{Xu,Wojek}.

We have also illustrated the expectation value of the pseudospin in the
vicinity of the $X$ point in Fig.\ref{SpinDirection}(b). The pseudospin
vector points the $x$-direction when $n^{\prime }=0$ in Eq.(\ref{HamilTCI}),
since then $\tau _{x}$ is a good quantum number. The pseudospin direction is
inverted at the oval, which is the interception of the two Dirac cones. When
$n^{\prime }\neq 0$, the magnitude of the pseudospin, $\mathbf{t}%
^{2}=t_{x}^{2}+t_{y}^{2}+t_{z}^{2}$, becomes quite small also around the
oval.

The fact that the magnitudes of the pure spin and pseudospin are quite small
around the oval leads to a strong entanglement of the spin and pseudospin
there, as we now argue. The Hamiltonian is described by the $4\times 4$
matrix, which results in the SU(4) group structure of the system. The SU(4)
group is decomposed into the pure spin and pseudospin parts and the
spin-pseudospin entangled part. The generators of the pure spin (pseudospin)
part are given by $\sigma _{i}$ ($\tau _{i}$) with $i=x,y,z$. On the other
hand, those of the spin-pseudospin entangled part are given by $\sigma
_{i}\tau _{j}$ with $i,j=x,y,z$, which compose the SU(2)$\otimes $SU(2)
group. The magnitude of the SU(4) spin is a constant and takes the same
value everywhere. Hence, the fact that the pure spin and pseudospin
components become quite small means that the spin-pseudospin entangled
components such as $\sigma _{z}\tau _{y}$ and $\sigma _{y}\tau _{z}$ become
large. The results implies a rich topological structure in the SU(4) space.

\section{Chern number and valley-Chern number}

The Chern number is obtained by the integration over the whole Brillouin
zone. We illustrate the Berry curvature $F\left( \mathbf{k}\right) $ of the
highest unoccupied state in Fig.\ref{SpinDirection}(c). The Berry curvature
is found to exhibit sharp peaks at the vortex centers of the spin rotation,
which correspond to the tips of the Dirac cones, and become zero away from
them. Hence, the Chern number is given by the sum of the contributions from
individual Dirac cones.\textbf{\ }Note that the Berry curvature at the $X$
point is exactly canceled out by the one from the other occupied band, and
does not contribute to the Chern number.

In the vicinity of the $\Lambda _{X}$ point, we obtain an analytic form for
the Berry curvature $F_{X}\left( \mathbf{k}\right) $ by using the low-energy
Hamiltonian (\ref{low}),%
\begin{equation}
F_{X}\left( \mathbf{k}\right) =\frac{\tilde{m}_{X}\tilde{v}_{1}\tilde{v}_{2}%
}{\left( \tilde{v}_{1}^{2}k_{x}^{2}+\tilde{v}_{2}^{2}k_{y}^{2}+\tilde{m}%
_{X}^{2}\right) ^{3/2}}.
\end{equation}%
The Chern number is explicitly calculated as%
\begin{equation}
\mathcal{C}_{X}=\frac{1}{2\pi }\int F\left( \mathbf{k}\right) d\mathbf{k}=%
\frac{1}{2}\text{sgn}(\tilde{m}_{X})=\frac{1}{2}\text{sgn}(m_{X}),
\label{ChernX}
\end{equation}%
which is associated with the Dirac cone at the $\Lambda _{X}$ point. The
similar formulas are derived for $\mathcal{C}_{X}^{\prime }$, $\mathcal{C}%
_{Y}$ and $\mathcal{C}_{Y}^{\prime }$ with the use of $m_{X}^{\prime }$, $%
m_{Y}$ and $m_{Y}^{\prime }$ for the Dirac cones at the $\Lambda
_{X}^{\prime }$, $\Lambda _{Y}$ and $\Lambda _{Y}^{\prime }$ points,
respectively.

At low energy there are four Dirac Hamiltonians such as (\ref{low}) and (\ref%
{lowY}), each of which describes a Dirac cone possessing a definite Chern
number depending on the sign of the Dirac mass. Hence\ there are four Chern
numbers. The genuine Chern number is their sum,%
\begin{equation}
\mathcal{C}=\mathcal{C}_{X}+\mathcal{C}_{X}^{\prime }+\mathcal{C}_{Y}+%
\mathcal{C}_{Y}^{\prime }.
\end{equation}%
This is a genuine topological number.

In addition, there are three valley-Chern numbers\cite{SPT}, which we may
take as
\begin{subequations}
\begin{align}
\mathcal{C}_{1}=& \mathcal{C}_{X}+\mathcal{C}_{X}^{\prime }-\mathcal{C}_{Y}-%
\mathcal{C}_{Y}^{\prime }, \\
\mathcal{C}_{2}=& \mathcal{C}_{X}-\mathcal{C}_{X}^{\prime }+\mathcal{C}_{Y}-%
\mathcal{C}_{Y}^{\prime }, \\
\mathcal{C}_{3}=& \mathcal{C}_{X}-\mathcal{C}_{X}^{\prime }-\mathcal{C}_{Y}+%
\mathcal{C}_{Y}^{\prime }.
\end{align}%
\end{subequations}
They are symmetry-protected topological numbers. The relevant symmetry is
the valley symmetry, which is the permutation symmetry of Dirac valleys.
This is a good symmetry near the Fermi level, since the system is described
by four Dirac Hamiltonians independent each other. However, at higher
energy, the system is described by the tight-binding Hamiltonian, containing
inter-valley hoppings, where there is no valley symmetry.

If we treat the four masses independently there are $16$ topological states
indexed by $(\mathcal{C},\mathcal{C}_{1},\mathcal{C}_{2},\mathcal{C}_{3})$.
However, when there are constraints on them, they read as follows:

(1) When we apply only the exchange field $(\Delta m_{X}=\Delta m_{Y}=0)$,
we find $\pm (2,0,0,0)$ with $\mathcal{C}_{X}=\mathcal{C}_{Y}=\mathcal{C}%
_{X}^{\prime }=\mathcal{C}_{Y}^{\prime }$.

(2) When we apply only the strain $(m=0)$, we find $\pm (0,0,2,0)$ with $%
\mathcal{C}_{X}=\mathcal{C}_{Y}=-\mathcal{C}_{X}^{\prime }=-\mathcal{C}%
_{Y}^{\prime }$ for $\Delta m_{X}\Delta m_{Y}>0$, and $\pm (0,0,0,2)$ with $%
\mathcal{C}_{X}=-\mathcal{C}_{Y}=-\mathcal{C}_{X}^{\prime }=\mathcal{C}%
_{Y}^{\prime }$ for $\Delta m_{X}\Delta m_{Y}<0$.

(3) When we apply both the exchange field and the strain to the crystal, we
find $\pm (1,-1,1,1)$ for $\Delta m_{X}>m>0$ and $m>\Delta m_{Y}>0$.

There are some other cases depending on $m$, $\Delta m_{X}$ and $\Delta
m_{Y} $. We have found that the Chern number may take values $2,1,0,-1,-2$.
Even if it is zero, the state is topological with respect to the
valley-Chern numbers.

\section{Optical absorption and elliptic dichroism}

An interesting experiment to probe and manipulate the valley degree of
freedom is to employ the optical absorption\cite%
{Yao08,Xiao,Li,EzawaOpt,Stille,Carbotte}. It is possible to excite massive
Dirac electrons by the right or left circularly polarized light, known as
circular dichroism. Originally, circular dichroism is proposed in honeycomb
systems, where the velocities of the Dirac cones are isotropic. On the other
hand they are anisotropic in the TCI surface. This leads to the elliptic
dichroism, where the optical absorptions are different between the right and
left elliptically polarized lights. Furthermore, the optical absorptions
depend crucially on the sign of the Dirac mass.

\subsection{Kubo formula}

We explore optical inter-band transitions from the state $|u_{\text{v}}(%
\widetilde{\mathbf{k}})\rangle $ in the valence band to the state $|u_{\text{%
c}}(\widetilde{\mathbf{k}})\rangle $ in the conduction band. The fundamental
transition is a transition from the highest occupied band to the lowest
unoccupied band (Fig.\ref{FigBand}). We inject a beam of elliptical
polarized light onto the TCI surface. The corresponding electromagnetic
potential is given by $\mathbf{A}(t)=(A_{x}\sin \omega t,A_{y}\cos \omega t)$%
. The electromagnetic potential is introduced into the Hamiltonian by way of
the minimal substitution, that is, by replacing the momentum $\widetilde{k}%
_{i}$ with the covariant momentum $P_{i}\equiv \widetilde{k}_{i}+eA_{i}$.
The resultant Hamiltonian simply reads $H\left( A\right) =H+\mathcal{P}%
_{x}A_{x}+\mathcal{P}_{y}A_{y}$, with
\begin{equation}
\mathcal{P}_{x}=\frac{\partial H}{\partial \widetilde{k}_{x}},\qquad
\mathcal{P}_{y}=\frac{\partial H}{\partial \widetilde{k}_{y}},
\end{equation}%
in the linear response theory.

The optical absorption is governed by the Fermi golden rule. Namely, the
imaginary part of the dielectric function arises due to inter-band
absorption, and is given by the Kubo formula. In the case of elliptical
polarized light it reads\cite{Yao08}%
\begin{align}
\varepsilon _{\theta }\left( \omega \right) =& \frac{\pi e^{2}}{\varepsilon
_{0}m_{e}^{2}\omega ^{2}}\sum_{i}\int_{BZ}\frac{d\widetilde{\mathbf{k}}}{%
\left( 2\pi \right) ^{2}}f(\widetilde{\mathbf{k}})\left\vert P_{\theta }(%
\widetilde{\mathbf{k}})\right\vert ^{2}  \notag \\
& \times \delta \left[ E_{\text{c}}(\widetilde{\mathbf{k}})-E_{\text{v}}(%
\widetilde{\mathbf{k}})-\omega \right] ,  \label{absorp}
\end{align}%
with the use of the optical matrix element $P_{\theta }(\widetilde{\mathbf{k}%
})$, where $E_{\text{c}}(\widetilde{\mathbf{k}})$ and $E_{\text{v}}(%
\widetilde{\mathbf{k}})$ are the energies of the conduction and valence
bands, while $f(\widetilde{\mathbf{k}})$ is the Fermi distribution function.
The coupling strength with optical fields is given by the optical matrix
element between the initial and final states in the photo-emission process%
\cite{Yao08,Xiao,Li,EzawaOpt},
\begin{equation}
P_{i}(\widetilde{\mathbf{k}})\equiv m_{0}\left\langle u_{\text{c}}(%
\widetilde{\mathbf{k}})\right\vert \frac{\partial H}{\partial \widetilde{k}%
_{i}}\left\vert u_{\text{v}}(\widetilde{\mathbf{k}})\right\rangle ,
\label{EqP}
\end{equation}%
which is the interband matrix element of the canonical momentum operator.
The optical matrix element for elliptically polarized light is
\begin{equation}
P_{\theta }(\widetilde{\mathbf{k}})=P_{x}(\widetilde{\mathbf{k}})\cos \theta
+iP_{y}(\widetilde{\mathbf{k}})\sin \theta ,
\end{equation}%
where $\theta $ is the ellipticity of the injected beam. We call it the
right polarized light for $0<\theta <\pi $ and the left one for $-\pi
<\theta <0$.

\begin{figure}[t]
\centerline{\includegraphics[width=0.4\textwidth]{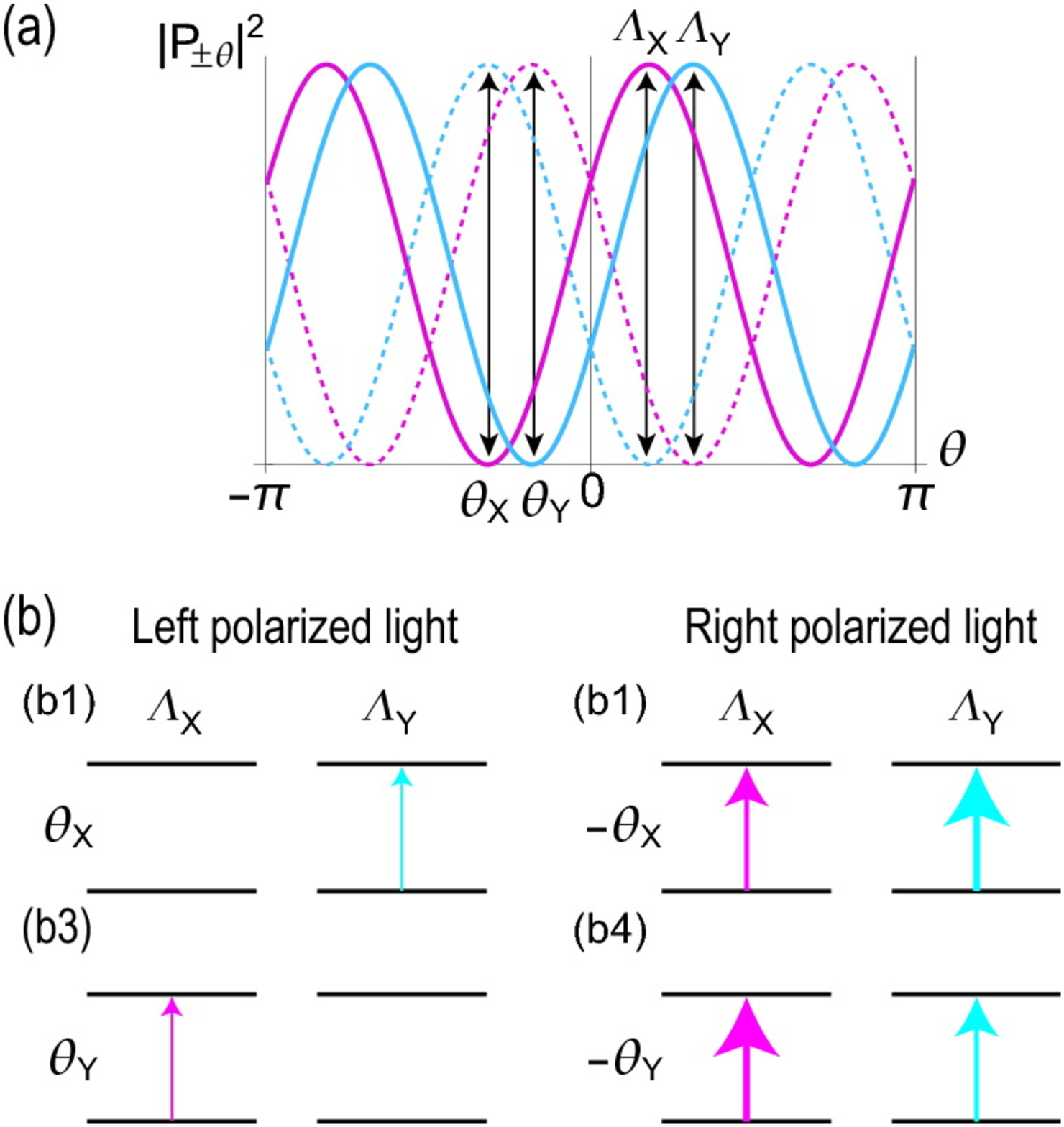}}
\caption{(Color online) (a) Optical matrix element $\left\vert P_{\pm
\protect\theta }\right\vert ^{2}$ at the $\Lambda _{X}$ and $\Lambda _{Y}$\
points with various ellipticity $\protect\theta $ [Eq.(\protect\ref{Sin})].
Red (blue) solid curves are optical absorption $\left\vert P_{\protect\theta %
}\right\vert ^{2}$ at the $X$ ($Y$) point, and dotted curves are for $%
\left\vert P_{-\protect\theta }\right\vert ^{2}$. (b) Illustration of
optical absorption $\left\vert P_{\protect\theta }\right\vert ^{2}$ at (b1) $%
\protect\theta =\protect\theta _{X}$, (b2) $\protect\theta =-\protect\theta %
_{X}$, (b3) $\protect\theta =\protect\theta _{Y}$, (b4) $\protect\theta =-%
\protect\theta _{Y}$. The magnitude of arrows indicates the magnitude of
optical absorption. We have assumed that all four masses have positive
values.}
\label{FigSin}
\end{figure}

\subsection{Optical absorption at the Dirac point}

We first investigate optical interband transitions from the valence-band
tops to the conduction band bottoms, i.e., at the Dirac point. By adjusting
the energy of light to the band edge, namely, at $\widetilde{\mathbf{k}}=0$,
\begin{equation}
\omega =E_{\text{c}}(0)-E_{\text{v}}(0)=2\left\vert \tilde{m}\right\vert ,
\end{equation}%
we find%
\begin{equation}
\varepsilon _{\theta }\left( 2\left\vert \tilde{m}\right\vert \right) =\frac{%
\pi e^{2}}{4\varepsilon _{0}m_{e}^{2}\tilde{m}^{2}}\left\vert P_{\theta
}\left( 0\right) \right\vert ^{2}
\end{equation}%
at each Dirac point, where $\tilde{m}$ can be any of $\tilde{m}_{X}$, $%
\tilde{m}_{X}^{\prime }$, $\tilde{m}_{Y}$, $\tilde{m}_{Y}^{\prime }$. It
follows that $\left\vert P_{\theta }\left( 0\right) \right\vert ^{2}$ can be
directly observed by optical absorption.

The wave functions $|u_{\text{v}}(\widetilde{\mathbf{k}})\rangle $ and $|u_{%
\text{c}}(\widetilde{\mathbf{k}})\rangle $ are obtained explicitly by
diagonalizing Eq.(\ref{HamilTCI}), and we have
\begin{equation}
P_{x}(0)=\tilde{v}_{1},\qquad P_{y}(0)=-i\tilde{v}_{2}\text{sgn}\left[ m_{X}%
\right] .
\end{equation}%
It is possible to derive an explicit form of $|P_{\theta }^{\pm
}(0)|_{\Lambda _{X}}^{2}$ at the $\Lambda _{X}$ point for arbitrary
ellipticity $\theta $ as%
\begin{equation}
\left\vert P_{\theta }(0)\right\vert _{\Lambda _{X}}^{2}=m_{0}^{2}\left(
\tilde{v}_{1}\cos \theta +\text{sgn}\left[ m_{X}\right] \tilde{v}_{2}\sin
\theta \right) ^{2}.  \label{SinXX}
\end{equation}%
Similar formulas follow at the other Dirac points. By introducing%
\begin{equation}
\tan \phi _{X}=\tilde{v}_{1}/\tilde{v}_{2},\qquad \tan \phi _{Y}=\tilde{v}%
_{2}/\tilde{v}_{1},
\end{equation}%
we rewrite them as
\begin{subequations}
\label{Sin}
\begin{align}
\left\vert P_{\theta }(0)\right\vert _{\Lambda _{X}}^{2}=& m_{0}^{2}\left(
\tilde{v}_{1}^{2}+\tilde{v}_{2}^{2}\right) \sin ^{2}\left( \phi _{X}+\text{%
sgn}\left[ m_{X}\right] \theta \right) , \\
\left\vert P_{\theta }(0)\right\vert _{\Lambda _{X}^{\prime }}^{2}=&
m_{0}^{2}\left( \tilde{v}_{1}^{2}+\tilde{v}_{2}^{2}\right) \sin ^{2}\left(
\phi _{X}+\text{sgn}\left[ m_{X}^{\prime }\right] \theta \right) ,
\end{align}%
and%
\begin{align}
\left\vert P_{\theta }(0)\right\vert _{\Lambda _{Y}}^{2}=& m_{0}^{2}\left(
\tilde{v}_{1}^{2}+\tilde{v}_{2}^{2}\right) \sin ^{2}\left( \phi _{Y}+\text{%
sgn}\left[ m_{Y}\right] \theta \right) ,  \label{SinY1} \\
\left\vert P_{\theta }(0)\right\vert _{\Lambda _{Y}^{\prime }}^{2}=&
m_{0}^{2}\left( \tilde{v}_{1}^{2}+\tilde{v}_{2}^{2}\right) \sin ^{2}\left(
\phi _{Y}+\text{sgn}\left[ m_{Y}^{\prime }\right] \theta \right) .
\label{SinY2}
\end{align}%
\end{subequations}
We note that
\begin{equation}
\phi _{X}=0.317\pi ,\qquad \phi _{Y}=0.183\pi  \label{EqD}
\end{equation}%
for $v_{1}=1.3$eV, $v_{2}=2.4$eV, and that%
\begin{equation}
\phi _{X}+\phi _{Y}=\frac{\pi }{2}\qquad (\text{mod}\pi ).  \label{phiXY}
\end{equation}%
There are four functions with the same amplitude in general: See Fig.\ref%
{FigSin}(a). The function (red solid curve) involving $|\sin \left( \phi
_{X}+\theta \right) |^{2}$ is the main one. The function (blue solid curve)
involving $|\sin \left( \phi _{Y}+\theta \right) |^{2}$ is constructed by
sifting it so that (\ref{phiXY}) holds. The other two functions (dotted
curves) are constructed by changing $\theta \rightarrow -\theta $.

For instance, when all masses are positive such as in the case of the
exchange effect, it follows that $\left\vert P_{\theta }(0)\right\vert
_{\Lambda _{X}}^{2}=\left\vert P_{\theta }(0)\right\vert _{\Lambda
_{X}^{\prime }}^{2}$, as is shown in the red solid lines in Fig.\ref{FigSin}%
(a). It also follows that $\left\vert P_{\theta }(0)\right\vert _{\Lambda
_{Y}}^{2}=\left\vert P_{\theta }(0)\right\vert _{\Lambda _{Y}^{\prime }}^{2}$%
, as is shown in blue solid curves in Fig.\ref{FigSin}(a).

For instance, when $m_{X}m_{X}^{\prime }<0$ and $m_{Y}m_{Y}^{\prime }<0$
such as in the case of the strain effect, it follows that $\left\vert
P_{\theta }(0)\right\vert _{\Lambda _{X}}^{2}=\left\vert P_{-\theta
}(0)\right\vert _{\Lambda _{X}^{\prime }}^{2}$ and $\left\vert P_{\theta
}(0)\right\vert _{\Lambda _{Y}}^{2}=\left\vert P_{-\theta }(0)\right\vert
_{\Lambda _{Y}^{\prime }}^{2}$. Thus, if $m_{X}>0$ and $m_{Y}>0$, they are
described by the same solid curves at the $\Lambda _{X}$ and $\Lambda _{Y}$
points but by the dotted curves at the $\Lambda _{X}^{\prime }$ and $\Lambda
_{Y}^{\prime }$ points in Fig.\ref{FigSin}(a).

A perfect elliptic dichroism is a phenomenon that only one-handed
elliptically polarized light is absorbed. It occurs at $\theta =-\phi _{X}$
for the function $|\sin \left( \phi _{X}+\theta \right) |^{2}$. At the same
point the function $|\sin \left( \phi _{Y}+\theta \right) |^{2}$ takes the
maximum value. More explicitly they occur as $\theta =\theta _{X}$ at the $%
\Lambda _{X}$ point and so on, with%
\begin{align}
\theta _{X}=& -\text{sgn}\left[ m_{X}\right] \phi _{X},\qquad \theta
_{X}^{\prime }=-\text{sgn}\left[ m_{X}^{\prime }\right] \phi _{X},  \notag \\
\theta _{Y}=& -\text{sgn}\left[ m_{Y}\right] \phi _{Y},\qquad \theta
_{X}^{\prime }=-\text{sgn}\left[ m_{Y}^{\prime }\right] \phi _{Y}.
\end{align}%
We give an example in Fig.\ref{FigSin}(a) when all the masses are positive,
where $\theta _{X}^{\prime }=\theta _{X}$ and $\theta _{X}^{\prime }=\theta
_{X}$.

\begin{figure}[t]
\centerline{\includegraphics[width=0.48\textwidth]{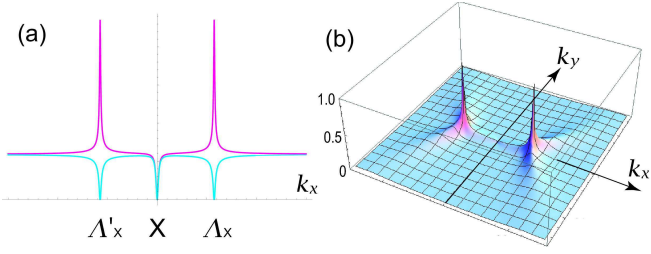}}
\caption{(Color online) (a) Optical matrix element $|P_{\protect\theta _{X}}(%
\widetilde{\mathbf{k}})|^{2}$ (red curve) and $|P_{-\protect\theta _{X}}(%
\widetilde{\mathbf{k}})|^{2}$ (blue curve) along the $\widetilde{k}_{x}$
axis, (b) $\widetilde{\mathbf{k}}$-resolved optical polarization $\protect%
\eta (\widetilde{\mathbf{k}})$. It has two sharp peaks at the $\Lambda _{X}$
and $\Lambda _{X}^{\prime }$ points. We have taken $\tilde{m}_{X}=\tilde{m}%
_{Y}=2$meV. }
\label{FigOpt}
\end{figure}

We have studied analytically the optical matrix element $|P_{\theta }(%
\widetilde{\mathbf{k}})|$ at the Dirac point. Next we investigate it away
from the Dirac point. An analytic solution of the optical matrix element of
right and left elliptically polarized light $|P_{\theta }(\widetilde{\mathbf{%
k}})|^{2}$\ is obtained from Eq.(\ref{HamilTCI}). However, the expression is
very complicated. We show the result in Fig.\ref{FigOpt} at $\theta =\theta
_{X}$, which shows the low-energy Dirac theory captures the essential
features. There are sharp peaks in optical absorption near the $\Lambda _{X}$
($\Lambda _{X}^{\prime }$) points. Fig.\ref{FigOpt}(a) shows the optical
matrix element $|P_{\theta _{X}}(\widetilde{\mathbf{k}})|^{2}$ and $%
|P_{-\theta _{X}}(\widetilde{\mathbf{k}})|^{2}$ along the $\widetilde{k}_{x}$
axis. We clearly see the difference between the right and left polarized
lights at the $\Lambda _{X}$ ($\Lambda _{X}^{\prime }$) point. There is
large optical absorption in right polarized light, while no optical
absorption in left polarized light. This is a dichroism caused by
elliptically polarized light, and the key feature of the elliptic dichroism.

\subsection{Optical absorption away from the Dirac point}

We proceed to drive the analytic expression of $|P_{\theta }(\widetilde{%
\mathbf{k}})|$ away from the $\Lambda _{X}$ point with the use of the the
low-energy Hamiltonian (\ref{low}) in (\ref{EqP}). It is straightforward to
find that
\begin{subequations}
\begin{align}
P_{x}(\widetilde{\mathbf{k}})& =\tilde{v}_{1}\frac{\tilde{v}_{1}\widetilde{k}%
_{x}\tilde{m}_{X}+i\tilde{v}_{2}\widetilde{k}_{y}\sqrt{\tilde{m}_{X}^{2}+%
\tilde{v}_{1}^{2}\widetilde{k}_{x}^{2}+\tilde{v}_{2}^{2}\widetilde{k}_{y}^{2}%
}}{\sqrt{\tilde{v}_{1}^{2}\widetilde{k}_{x}^{2}+\tilde{v}_{2}^{2}\widetilde{k%
}_{y}^{2}}\sqrt{\tilde{m}_{X}^{2}+\tilde{v}_{1}^{2}\widetilde{k}_{x}^{2}+%
\tilde{v}_{2}^{2}\widetilde{k}_{y}^{2}}},  \label{EqA} \\
P_{y}(\widetilde{\mathbf{k}})& =\tilde{v}_{2}\frac{\tilde{v}_{2}\widetilde{k}%
_{y}\tilde{m}_{X}-i\tilde{v}_{1}\widetilde{k}_{x}\sqrt{\tilde{m}_{X}^{2}+%
\tilde{v}_{1}^{2}\widetilde{k}_{x}^{2}+\tilde{v}_{2}^{2}\widetilde{k}_{y}^{2}%
}}{\sqrt{\tilde{v}_{1}^{2}\widetilde{k}_{x}^{2}+\tilde{v}_{2}^{2}\widetilde{k%
}_{y}^{2}}\sqrt{\tilde{m}_{X}^{2}+\tilde{v}_{1}^{2}\widetilde{k}_{x}^{2}+%
\tilde{v}_{2}^{2}\widetilde{k}_{y}^{2}}},  \label{EqB}
\end{align}%
since $\mathcal{P}_{x}=\tilde{v}_{1}\sigma _{x}$ and $\mathcal{P}_{y}=\tilde{%
v}_{2}\sigma _{y}$. At $\theta =\theta _{X}$, it yields a simple form,
\end{subequations}
\begin{equation}
|P_{\theta _{X}}(\widetilde{\mathbf{k}})|^{2}=m_{0}^{2}\tilde{v}_{1}\tilde{v}%
_{2}\frac{\left( \pm \tilde{m}_{X}+\sqrt{\tilde{m}_{X}^{2}+\tilde{v}_{1}^{2}%
\widetilde{k}_{x}^{2}+\tilde{v}_{2}^{2}\widetilde{k}_{y}^{2}}\right) ^{2}}{%
\tilde{m}_{X}^{2}+\tilde{v}_{1}^{2}\widetilde{k}_{x}^{2}+\tilde{v}_{2}^{2}%
\widetilde{k}_{y}^{2}}.  \label{P}
\end{equation}%
We derive the same formula away from the $\Lambda _{X}^{\prime }$ point just
replacing $\tilde{m}_{X}$ with $\tilde{m}_{X}^{\prime }$. Similar formulas
are derived also with respect to the $\Lambda _{Y}$ and $\Lambda
_{Y}^{\prime }$ points.

\begin{figure}[t]
\centerline{\includegraphics[width=0.3\textwidth]{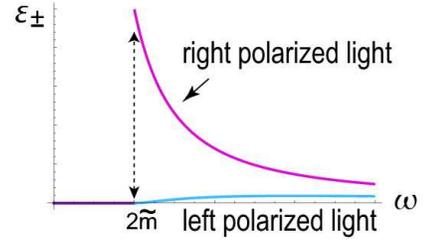}}
\caption{(Color online) Imaginary part of dielectric function $\protect%
\varepsilon _{\pm }\left( \protect\omega \right) $ due to interband
absorptions at $\protect\theta =\protect\theta _{X}$: See Eq.(\protect\ref{W}%
). A clear difference is observed between the right and left polarized
lights. There is almost no optical absorption for left polarized light for $%
\protect\omega >2\tilde{m}$. We have taken $\tilde{m}>0$ for definateness.}
\label{FigW}
\end{figure}

Representing (\ref{P}) in terms of the energy (\ref{ELow}), we obtain
\begin{equation}
|P_{\theta }(\widetilde{\mathbf{k}})|^{2}=m_{0}^{2}\tilde{v}_{1}\tilde{v}_{2}%
\frac{(\pm \tilde{m}+E_{\text{v}}(\widetilde{\mathbf{k}}))^{2}}{\left[ E_{%
\text{v}}(\widetilde{\mathbf{k}})\right] ^{2}},  \label{PE}
\end{equation}%
at $\theta =\theta _{X(Y)}$ with the use of $\tilde{m}=\tilde{m}_{X(Y)}$,
and $\theta =\theta _{X(Y)}^{\prime }$ with the use of $\tilde{m}=\tilde{m}%
_{X(Y)}^{\prime }$, where we have used the relation $\varepsilon _{\text{v}}(%
\widetilde{\mathbf{k}})=-\varepsilon _{\text{c}}(\widetilde{\mathbf{k}})$
required by the electron-hole symmetry of the energy spectrum.

We substitute (\ref{PE}) to (\ref{absorp}), and use the density of state%
\begin{equation}
\rho \left( E\right) =\frac{\left\vert E\right\vert }{2\pi \tilde{v}_{1}%
\tilde{v}_{2}}\Theta \left( E-2|\tilde{m}|\right)
\end{equation}%
with the step function $\Theta \left( x\right) =1$ for $x>0$ and $\Theta
\left( x\right) =0$ for $x<0$, to find%
\begin{equation}
\varepsilon _{\pm }\left( \omega \right) =\frac{e^{2}m_{0}^{2}}{2\varepsilon
_{0}m_{e}^{2}\omega }\frac{\left( \pm \tilde{m}+\omega /2\right) ^{2}}{%
\left( \hbar \omega /2\right) ^{2}}\Theta \left( \omega -2|\tilde{m}|\right)
.  \label{W}
\end{equation}%
Hence there is no optical absorption for%
\begin{equation}
\omega =\mp 2\tilde{m}>0.  \label{EqC}
\end{equation}%
We show the optical absorption (\ref{W}) in Fig.\ref{FigW}. A clear
difference is observed between the right- and left- polarized lights. There
is almost no optical absorption for left polarized light for $\omega >2|%
\tilde{m}|$. Here, $\tilde{m}$ stands for any of $\tilde{m}_{X}$, $\tilde{m}%
_{X}^{\prime }$, and $\tilde{m}_{Y}^{\prime }$.

A perfect elliptic dichroism follows that $|P_{\theta _{X}}\left( 0\right)
|^{2}=0$ if $m_{X}>0$, while $|P_{-\theta _{X}}\left( 0\right) |^{2}=0$ if $%
m_{X}<0$. The anisotropy of the Dirac cone is determined by measuring the
ellipticity angle $\theta _{X}$ of the injected beam: See Fig.\ref{FigSin}.
We would expect $\theta _{X}=0.317\pi $ as in (\ref{EqD}). We can also
determine the band gap by measuring the energy where the optical absorption
becomes nonzero (\ref{W}): See Fig.\ref{FigW}. The role of the right- and
left-polarized light is inverted when the sign of the mass term is negative.
Thus we can determine the sign of the mass term by the elliptic dichroism
even when the magnitude of the mass term is very small.

\subsection{Optical polarization}

We next investigate the $k$-resolved optical polarization $\eta _{\theta }(%
\widetilde{\mathbf{k}})$, which is given by\cite{Yao08,Xiao,Li,EzawaOpt}%
\begin{equation}
\eta _{\theta }(\widetilde{\mathbf{k}})=\frac{|P_{\theta }(\widetilde{%
\mathbf{k}})|^{2}-|P_{-\theta }(\widetilde{\mathbf{k}})|^{2}}{|P_{\theta }(%
\widetilde{\mathbf{k}})|^{2}+|P_{-\theta }(\widetilde{\mathbf{k}})|^{2}},
\end{equation}%
which we show in Fig.\ref{FigOpt}(b). This quantity is the difference
between the absorption of the left- and right-handed lights ($\pm \theta $),
normalized by the total absorption, around the $\Lambda _{X}$ point. Optical
polarizations are perfectly polarized at the $\Lambda _{X}$ and $\Lambda
_{X}^{\prime }$ points ($\widetilde{\mathbf{k}}=0$). Namely, the selection
rule holds exactly at the $\Lambda _{X}$ and $\Lambda _{X}^{\prime }$
points. Then, $|\eta _{\theta }(\widetilde{\mathbf{k}})|$ rapidly decreases
to $0$ as $\left\vert \widetilde{\mathbf{k}}\right\vert $ increases.

\subsection{Valley-selective optical pumping}

An interesting valleytronics application of the elliptic dichroism would
read as follows. Let us adjust the ellipticity of light at $\theta =\theta
_{X}$ so that the optical absorption near the $\Lambda _{X}$\ point does not
occur [Fig\ref{FigSin}(b1)]. Then the optical absorption is not zero at the $%
\Lambda _{Y}$ point. Namely, we can selectively excite electrons at the $%
\Lambda _{Y}$\ point by left polarized light. It is a valley-selective
optical pumping. In the same way, by adjusting $\theta =\theta _{Y}$, we can
selectively excite electrons at the $\Lambda _{X}$ point by left polarized
light [Fig\ref{FigSin}(b3)]. The valley-selective optical pumping is
possible since the anisotropy of Dirac cones at $\Lambda _{X}$ and $\Lambda
_{Y}$ points are different. If the Dirac cones were isotropic, we could not
differentiate the Dirac cones at $\Lambda _{X}$ and $\Lambda _{Y}$ points
since they have the same chirality. This will pave a new way to
valleytronics in the TCI.

\section{Conclusions}

We have investigated the optical absorption on the TCI surface when gaps are
given to surface Dirac cones. First, the chiralities of all four Dirac cones
are identical, which can be verified by studying the spin direction.
Nevertheless, it is possible to make a selective excitation between the $%
\Lambda _{X}$ ($\Lambda _{X}^{\prime }$) point and the $\Lambda _{Y}$ ($%
\Lambda _{Y}^{\prime }$) point, because the Dirac cones are anisotropic,
where $\tilde{v}_{2}/\tilde{v}_{1}=0.65$. Furthermore, it is also possible
to make a selective excitation between the $\Lambda _{X}$ and $\Lambda
_{X}^{\prime }$ points when the Dirac masses $\tilde{m}_{X}$ and $\tilde{m}%
_{X}^{\prime }$ have the opposite signs. Namely, by tuning the ellipticity
of the polarized light, we can realize a perfect elliptic dichroism, where
only electrons at one valley are excited. Our results will pave a new road
toward valleytronics based on the TCI.

\label{SecConclusion}

I am very much grateful to N. Nagaosa, Y. Ando, L. Fu and T. H. Hsieh for
many helpful discussions on the subject. This work was supported in part by
Grants-in-Aid for Scientific Research from the Ministry of Education,
Science, Sports and Culture No. 22740196.

\end{document}